\documentclass[conference]{IEEEtran}
\IEEEoverridecommandlockouts
\usepackage{cite}
\usepackage{amsmath,amssymb,amsfonts}
\usepackage{algorithmic}
\usepackage{graphicx}
\usepackage{wrapfig}
\usepackage{pgf-pie}
\usepackage{textcomp}
\usepackage{xcolor}
\usepackage{listings}
\usepackage{multicol}
\usepackage[binary-units]{siunitx}
\usepackage[inline]{enumitem}
\usepackage{multirow}
\usepackage{hyperref}
\newcolumntype{?}{!{\vrule width 1pt}}
\usepackage{listings}
\usepackage{pifont}
\usepackage{csquotes}

\lstdefinelanguage{text}{}

\lstdefinestyle{styledScriptText}{
  language=text,
  columns=flexible,
  basicstyle=\scriptsize\ttfamily,
  keywordstyle=\color{blue}\bfseries,
  showstringspaces=false,
  breaklines=true,
  commentstyle=\color{forestgreen}\ttfamily,
  extendedchars=true, 
  framesep=4pt, 
  numbers=none, 
  alsoletter={0123456789.}
}

\lstdefinestyle{styledSmallText}{
  language=text,
  columns=flexible,
  basicstyle=\small\ttfamily,
  keywordstyle=\color{blue}\bfseries,
  showstringspaces=false,
  breaklines=true,
  commentstyle=\color{forestgreen}\ttfamily,
  extendedchars=true, 
  framesep=4pt, 
  numbers=none, 
  alsoletter={0123456789.}
}

\lstdefinestyle{styledFootnoteText}{
  language=text,
  columns=flexible,
  basicstyle=\footnotesize\ttfamily,
  keywordstyle=\color{blue}\bfseries,
  showstringspaces=false,
  breaklines=true,
  commentstyle=\color{forestgreen}\ttfamily,
  extendedchars=true, 
  framesep=4pt, 
  numbers=none, 
  alsoletter={0123456789.}
}

\def\BibTeX{{\rm B\kern-.05em{\sc i\kern-.025em b}\kern-.08em
    T\kern-.1667em\lower.7ex\hbox{E}\kern-.125emX}}
\begin{document}

\title{From Monolith to Microservices \\ Static and Dynamic Analysis Comparison}

\author{Bernardo~Andrade,
        Samuel~Santos and
        António~Rito~Silva
\IEEEcompsocitemizethanks{\IEEEcompsocthanksitem D. Faustino, N. Gonçalves and A. Rito Silva are with DPSS - INESC-ID, Instituto Superior Técnico, University of Lisbon, Lisbon, Portugal.
E-mail: \{bernardo.andrade, samuel.c.santos, rito.silva\}@tecnico.ulisboa.pt
}
}

\maketitle
\begin{abstract}
One of the most challenging problems in the migration of a monolith to a microservices architecture is the identification of the microservices boundaries. Several approaches have been recently proposed for the automatic identification of microservices, which, even though following the same basic steps, diverge on how data of the monolith system is collected and analysed. In this paper, we compare the decompositions generated for two monolith systems into a set of candidate microservices, when static and dynamic analysis data collection techniques are used. The decompositions are generated using a combination of similarity measures and are evaluated according to a complexity metric to answer the following research question: which collection of monolith data, static or dynamic analysis, allows to generate better decompositions? As result of the analysis we conclude that neither of the analysis techniques, static nor dynamic, outperforms the other, but the dynamic collection of data requires more effort.
\end{abstract}

\begin{IEEEkeywords}
Microservices, Software Evolution, Static Analysis, Dynamic Analysis, Software Architecture.
\end{IEEEkeywords}

\section{Introduction}

Microservices~\cite{Fowler14} have become main stream in the development of large scale and complex systems when companies, like Amazon and Netflix~\cite{Hanlon06}, faced constraints on their systems evolution, due to the coupling resulting from the use of a large domain model maintained in a shared database. However, the adoption of this architectural style is not free of problems~\cite{Taibi18}, where the identification of microservices boundaries is one of the most challenging, because a wrong cut results on the need to refactor between distributed services, which impacts on the services interfaces, and cannot have the support of integrated development environments. 

The microservices boundaries identification has been addressed by research, e.g.~\cite{Amiri18,Fritzsch18,Mazlami17,Tyszberowicz18,Jin19}, in the context of the migration of monolith systems to a microservices architecture. Some approaches take advantage of the monolith's codebase and runtime behavior to collect data, analyse it, and propose a decomposition of the monolith. Although each of the approaches use different techniques, they follow the same basic steps: (1) Collection: collect data from the monolith system; (2) Decomposition: define a decomposition by applying a similarity measure and an aggregation algorithm, like a clustering algorithm, to the data collected in the first step; (3) Analysis: evaluate the quality of the generated decomposition using a set of metrics. 

However, the approaches differ on the techniques applied in each one of the steps. In terms of the collection of data, they differ in whether it is collected from the monolith using static analysis of the code~\cite{Tyszberowicz18}, or if they observe the monolith execution behavior~\cite{Jin19}.

In this paper we analyse two monolith systems to study whether these techniques provide significant differences when identifying candidate decompositions. The analysis framework is built on top of what is considered, by the gray literature, as one of the main difficulties on the identification of microservices boundaries in monolith systems: the transactional contexts~\cite[Chapter~5]{Ford17}. Transactional contexts generate a coupling between domain entities accessed in the context of the same transaction, due to the complexity of decomposing a transactional behavior into several distributed transactions, problem known as the forgetting of the CAP theorem~\cite{Carrasco18}. Therefore, the decomposition to a microservices architecture should minimize the number of distributed transactions implementing a functionality, i.e., minimize the cost of redesigning the functionality in the microservices architecture.

Considering this analysis framework, we address the following research question: which collection of monolith behavior data, static or dynamic analysis, allows to generate better decompositions?

In this section we defined the context of our work. The next section formalizes our analysis framework. Section~\ref{section:collection} describes the overall process of automatic identification of candidate microservices and the use of the static and dynamic data collection techniques in particular. In the evaluation, Section~\ref{section:evaluation}, the analysis framework is applied to 2 systems in order to answer the research question. Section~\ref{section:related} presents related work and Section~\ref{section:discussion} discusses the outcomes of this work. Finally, Section~\ref{section:conclusions} presents the conclusions.

\section{Similarity Measures and Complexity Metric}
\label{formalization}

A monolith is defined by its set of functionalities which execute in atomic transactional contexts and, due to the migration to the microservices architecture, have to be decoupled into a set of distributed transactions, each one executing in the context of a microservice. 

Therefore, a monolith is defined as a triple $(F, E, G)$, where $F$ defines its set of functionalities, $E$ the set of domain entities, and $G$ a set of call graphs, one for each monolith functionality. A call graph is defined as a tuple $(A, P)$, where $A = E \times M$ is a set of read and write of accesses to domain entities ($M = \{r, w\}$), and $P = A \times A$ a precedence relation between elements of $A$ such that each access has zero or one immediate predecessors, $\forall_{a \in A} \#\{(a_1,a_2) \in P: a_1 = a \} \leq 1$, and there are no circularities, $\forall_{(a_1,a_2) \in P_T} (a_2,a_1) \notin P_T$, where $P_T$ is the transitive closure of $P$. The precedence relation represents the sequences of accesses associated with a functionality.

\subsection{Similarity Measures}

The definition of similarity measures establishes the distance between domain entities. Domain entities that are closer, according to a particular similarity measure, should be in the same microservice. Therefore, since we are interested in reducing the number of distributed transactions a functionality is decomposed in, we intend to define as close the domain entities that are accessed by the same functionalities.

The access similarity measure measures the distance between two domain entities, $e_1,e_2 \in E$, as:

\begin{equation*}
sm_{access}(e_1,e_2) = \frac{\#(funct(e_1) \cap funct(e_2))}{\#funct(e_1)}
\end{equation*}

where $funct(e)$ denotes the set of functionalities in the monolith whose call graph has a read or write access to $e$. This measure takes a value in the interval 0..1. When all the functionalities that access $e_1$ also access $e_2$ then it takes the value 1.

Since the cost of reading and writing is different in the context of distributed transactions, because writes introduce new intermediate states in the decomposition of a functionality, the next two similarity measures distinguish read from write accesses in order to reduce the number of write distributed transactions:
\begin{equation*}
sm_{read}(e_1,e_2) = \frac{\#(funct(e_1,r) \cap funct(e_2,r))}{\#funct(e_1,r)}
\end{equation*}
\begin{equation*}
sm_{write}(e_1,e_2) = \frac{\#(funct(e_1,w) \cap funct(e_2,w))}{\#funct(e_1,w)}
\end{equation*}
where $funct(e,m)$ denotes the set of functionalities in the monolith whose call graph has an access according to mode $m$, read or write, respectively. These two measures tend to include in the same microservice, domain entities that are read or written together, respectively.

Finally, another similarity measure that is found in the literature groups domain entities that are frequently accessed in sequence, in order to reduce the number of remote invocations between microservices, i.e., the domain entities that are frequently accessed in sequence should be in the same microservice. Therefore, the sequence similarity measure is defined:
\begin{equation*}
sm_{sequence}(e_1,e_2) = \frac{sumPairs(e_1,e_2)}{maxPairs}
\end{equation*}
where $sumPairs(e_1,e_2) = \sum_{f \in F} \#\{(a_i,a_j) \in G_f.P: (a_i.e = e_1 \land a_j.e = e_2) \lor (a_i.e = e_2 \land a_j.e = e_1)\})$, where $G_f.P$ is the precedence relation for functionality $f$, is the number of consecutive accesses of $e_1$ and $e_2$, and $maxPairs = max_{e_i,e_j\in E} (sumPairs(e_i,e_j))$ is the max number of consecutive accesses for two domain entities in the monolith.

\subsection{Complexity Metric}

A decomposition of a monolith is a partition of its domain entities set, where each element is included in exactly one subset, a cluster, and a partition of the call graph of each one of its functionalities. Therefore, given the call graph $G_f$ of a functionality $f$, and a decomposition $D \subseteq 2^E$, the partition call graph of a functionality $partition(G_f, D) = (LT, RI)$ is defined by a set of local transactions $LT$ and a set of remote invocations $RI$, where each local transaction
\begin{enumerate}[label=(\roman*)]
    \item is a subgraph of the functionality call graph, $\forall_{lt \in LT}: lt.A \subseteq G_f.A \land lt.P \subseteq G_f.P$;
    \item contains only accesses in a single cluster of the domain entities decomposition, $\forall_{lt \in LT} \exists_{c in D}: lt.A.e \subseteq c$;
    \item contains all consecutive accesses in the same cluster, $\forall_{a_i \in lt.A, a_j \in G_f.A}: ((a_i.e.c = a_j.e.c \land (a_i,a_j) \in G_f.P) \implies (a_i,a_j) \in lt.P) \lor ((a_i.e.c = a_j.e.c \land (a_j,a_i) \in G_f.P) \implies (a_j,a_i) \in lt.P)$.
\end{enumerate}

From the definition of local transaction, results the definition of remote invocations, which are the elements in the precedence relation that belong to different clusters, $RI = \{(a_i,a_j) \in G_f.P: a_i.e.c \neq a_j.e.c\}$. Note that, in these definitions, we use the dot notation to refer to elements of a composite or one of its properties, e.g., in $a_j.e.c$, $.e$ denotes the domain entity in the access, and $.c$ the cluster the domain entity belongs to.

The complexity for a functionality migration, in the context of a decomposition, is the effort required in the functionality redesign, because its transactional behavior is split into several distributed transactions, which introduce intermediate states due to the lack of isolation. Therefore, the following aspects have impact on the functionality migration redesign effort:
\begin{itemize}
    \item The number of local transactions, because each local transaction may introduce an intermediate state;
    \item The number of other functionalities that read domain entities written by the functionality, because it adds the need to consider the intermediate states between the execution of the different local transactions;
    \item The number of other functionalities that write domain entities read by the functionality, because the functionality redesign has to consider the different states these domain entities can be.
\end{itemize}

This complexity is associated with the cognitive load that the software developer has to address when redesigning a functionality. Therefore, the complexity metric is defined in terms of the functionality redesign.

\begin{equation*}
complexity(f,D) = \sum_{lt \in partition(G_f,D)} complexity(lt,D)
\end{equation*}

The complexity of a functionality is the sum of the complexities of its local transactions.
\begin{equation*}
\begin{split}
complexity(lt,D) = \#\cup_{a_i \in prune(lt)} \\
\{f_i \neq lt.f: dist(f_i,D) \land a_i^{-1} \in prune(f_i,D))\}
\end{split}
\end{equation*}

The complexity of a local transaction is the number of other distributed functionalities that read, or write, domain entities, written, or read, respectively by the local transaction. The auxiliary function $dist$ identifies distributed functionalities, given the decomposition; $a_i^{-1}$ denotes the inverse access, e.g. $(e_1,r)^{-1} = (e_i,w)$; and $prune$ denotes the relevant accesses inside a local transaction, by removing repeated accesses of the same mode to a domain entity. If both read and write accesses occur inside the same local transaction, they are both considered if the read occurs before the write. Otherwise, only the write access is considered.
These are the only accesses that have impact outside the local transaction.

\section{Monolith Microservices Identification}
\label{section:collection}

The different approaches to the migration of monoliths to microservices architectures apply, in the \textit{Collection} step, either static or dynamic techniques, but there is no evidence in the literature on whether one of them subsumes the other, whether they are equivalent, or even whether they are complementary. Therefore, we collected data using both techniques in order to address this open problem.

Data was collected from two monolith systems, LdoD\footnote{\url{https://github.com/socialsoftware/edition}} and Blended Workflow (BW)\footnote{\url{https://github.com/socialsoftware/blended-workflow}}, that are implemented using the Model-View-Controller architectural style, where the controllers process input events by triggering transactional changes in the model, thus, corresponding to monolith functionalities. The monolith is designed considering its controllers as transactions that manipulate a persistent model of domain entities. Our collection tool was developed to cope with the Spring-Boot\footnote{\url{https://spring.io/projects/spring-boot}} framework and the Fénix Framework\footnote{\url{https://fenix-framework.github.io/}} Object-Relational Mapper (ORM).

As result of the collection, the functionalities accesses are stored in JSON format. It consists in a mapping between functionality names and functionality objects, where each object has a traces field that consists in a list of trace objects. Each trace is characterized by a unique identifier and a (compressed) list of accesses observed for a specific functionality execution. An $Access$ is composed by the numeric identifier of the domain entity and the access type, either read or write.

During the \textit{Decomposition} step of the migration process, our tool uses hierarchical clustering (Python SciPy\footnote{\url{https://docs.scipy.org/doc/}}) to process the collected data and, according to the 4 similarity measures, generate a dendrogram of the domain entities. The generated dendrogram can be cut in order to produce different decompositions, given the number of clusters. Our decomposition tool supports different combinations of similarity measures, for instance, it is possible to generate a decomposition with the following weights (30\% access, 30\% read, 20\% write, 20\% sequence).

For the \textit{Analysis} step our tool generates multiple decompositions, by varying the similarity measures weights and the number of clusters, and compare them according to the complexity metric. Additionally, two different decompositions of the same system can be compared using the MoJoFM~\cite{mojoFM} distance metric, which will be use to compare the decompositions generated using statically and dynamically collected data. 

MoJoFM is a distance measure between two architectures expressed as a percentage. This measure is based on two key operations used to transform one decomposition into another: moves (Move) of entities between clusters, and merges (Join) of clusters. Given two decompositions, A and B, MoJoFM is defined as:
\begin{equation*}
MoJoFM (A,B) = (1 - \frac{mno(A,B)}{max(mno(\forall A,B))}) \times 100\%
\end{equation*}
where $mno(A, B)$ is the minimum number of Move and Join operations needed to transform A into B and
$max(mno(\forall A,B))$ is the number of Move and Join operations needed to transform the most distant decomposition into B.

\subsection{Data Collection Tools}

Two data collection tools were developed.
Spoon~\cite{pawlak2016spoon} is a static code analysis tool that provides an introspection API that allows to parse and analyse a Java codebase by simply giving its folders as input. It was customized to be applied to identify Spring-Boot controllers and persistent domain entities implemented using the FenixFramework ORM.

The dynamic data collection is done in a running instance of the monolith under analysis using~Kieker~\cite{kiekerTechnicalReport}. The monolith systems were instrumented using AspectJ\footnote{The Eclipse Foundation (2011). The AspectJ Project. \url{http://www.eclipse.org/aspectj/}} to intercept calls to the FenixFramework's data access methods, the ones responsible for manipulating the respective entity's persistent state.

\subsection{Monolith Monitoring}

While for the static data collection it was enough to run the customized Spoon tool on the monolith codebases, for the dynamic data collection three different monolith monitoring strategies were followed: in production, through functional testing, and by simulation.

Regarding the LdoD system, it was monitored in three different environments: production, functional testing and simulation. The production monitoring lasted 3 weeks and a total of \SI{490}{\giga\byte} worth of data was collected. Throughout this period, a tight supervision was necessary to oversee the impact the monitoring had on the performance of the system's functionalities. Since the server hosting the application had a small free disk space (around \SI{20}{\giga\byte}) and a massive drop in performance was observed if it was full, it was mandatory to collect the generated logs from time to time (2-3 days) to not harm the user experience and to gather fresh logs instead of discarding them. 

Analyzing the collected data presented in Table~\ref{tab:coveredControllersAndCoveredEntities}, only 44\% of the controllers were exercised in production, when compared with the total number of controllers identified by the static analysis. Therefore, further processing and evaluation of this data were abdicated due to the substantial effort required to process it and the relatively little coverage. 
Concerning functional testing, it was achieved by running a suite of 200 integration tests (4.207 lines of code) that exercised 96\% of the controllers and 82\% of the domain entities, generating a few megabytes ($<$\SI{200}{\mega\byte}) of data, while the instruction coverage, reported by JaCoCo\footnote{https://github.com/jacoco/jacoco}, was 72\% for domain entities and 82\% for controllers. The reduced size of the collected data is explained by the usage of small subset of the original database's data and so, the traces associated with the execution of functionalities were much shorter. 
Finally, an expert of the system simulated, during one hour, the use of functionalities, using a database with a minimal set of data, and \SI{200}{\mega\byte} of data was collected and 84\% of the controllers and 80\% of the domain entities were exercised.

In what concerns the BW system, it was only simulated by an expert during an hour and 86\% of entities and 68\% of controllers were exercised. In this case, the reduced number of exercised controllers is justified by the deprecation of several controllers that are not reachable through the user interface.

\begin{table}[t]
\centering
\caption{Coverage of dynamically collected data}
    \begin{tabular}{c|c|c|c|c|}
        \cline{2-5}
        & \multicolumn{3}{c|}{LdoD}& BW \\ \cline{2-5} 
        & Prod & Tests & Sim & Sim \\ \hline
        \multicolumn{1}{|c|}{Coverage Controllers (\%)} & 44 & 96 & 84 & 68 \\ \hline
        \multicolumn{1}{|c|}{Coverage Entities (\%)} & 79 & 82 & 80 & 86 \\ \hline
    \end{tabular}
\label{tab:coveredControllersAndCoveredEntities}
    \vspace{-2mm}
\end{table}

\subsection{Static vs Dynamic Data Collection}

The process of data collection obviously differ in the coverage of controllers between static and dynamic collection, and they also differ on the identification of the domain entities that each controller accesses.

\begin{table}[t]
    \centering
    \caption{Compare Collected Data - Average of identified entities per controller}
    \begin{tabular}{l|c|c?c|c?c|c|}
        \cline{2-7}
        & \multicolumn{4}{c?}{LdoD} & \multicolumn{2}{c|}{BW} \\ \cline{2-7} 
        & Static & Tests & Static & Sim & Static & Sim \\ \hline
        \multicolumn{1}{|l|}{AVG(Cov. E/C)} & 95\%  & 71\% & 91\% & 77\% & 93\% & 78\% \\ \hline
    \end{tabular}
    \label{tab:coverageControllersAndCoveredEntitiesPerController}
    \vspace{-2mm}
\end{table}

Table~\ref{tab:coverageControllersAndCoveredEntitiesPerController} presents the percentage (average) of domain entities that each controller accesses when comparing the different data collection strategies. For instance, in LdoD, static analysis identifies 95\% of the domain entities, when compared with the identified through tests, while tests identify 71\% of the domain entities, when compared with the identified through static analysis.

Therefore, we can observe that, for the coverage of the accesses to domain entities in the context of the controllers, in some cases, dynamic analysis can identify accesses to domain entities, in the context of a controller execution, that the static collection does not, due to late binding. This is one of the limitations of the static analysis that may not be able to statically infer the type of a domain entity, in the case of polymorphic inheritance. The opposite also occurs, static analysis can identify accesses to domain entities that dynamic analysis cannot, because depending on the inputs provided to controllers and data available in the database, some of the domain entities may not be accessed, both in tests and simulation.

\section{Evaluation}
\label{section:evaluation}

The goal of evaluation is to assess which technique, static or dynamic, provides the best results. First we evaluate whether the use of static or dynamic analysis allows to identify a combination of similarity measures that provides better decompositions, in terms of complexity. Then, we assess whether the dynamic analysis produces significantly different decompositions, when compared to the ones statically generated and with a source of truth. 

In both analysis, the \textit{Decomposition} step is going to be applied to the data collected, statically and dynamically. Therefore, 
several dendrograms are produced, by varying the weights of the four existing similarity measures - Access (A), Write (W), Read (R) and Sequence (S) - in intervals of 10 in a scale of 0 to 100. For instance (40, 20, 20, 20) represents a combination of similarity measures where a dendrogram is generated using hierarchical clustering for the 40\% access, 20\% write, 20\% read, and 20\% sequence.

Then several cuts are performed on each dendrogram. Each cut results in a candidate decomposition of the monolith with a specific number of clusters, varying from 3 to 10. For each generated decomposition, the values for the complexity metric are calculated. The complexity metric value had to be normalized in order to compare them among the two monoliths, since they depend on the number of functionalities of each monolith. 
The uniform complexity of a given decomposition $d$ of a monolith is calculated by dividing the complexity of $d$ by the $maxComplexity$. The $maxComplexity$ value is determined by calculating the complexity of a decomposition of the monolith where each cluster has a single domain entity. Therefore, the uniform complexity of any monolith decomposition is a value in the interval 0 to 1.

Therefore, in the experiments, we calculate, for each system, the uniform complexity of each decomposition generated by the combination of the 4 similarity measures, each varying in intervals of 10 and their sum being 100, and the number of clusters (N), between 3 and 10. 

\subsection{Complexity and Similarity Measures Correlation}

To assess the correlation between the complexity metric, the weights given to each similarity measure, and the number of clusters, a linear regression model was employed using the Ordinary Least Squares method, as given by:
\begin{equation*}
    \begin{split}
    \label{formula:simMeasures_complexity_regression}
        uComplexity (d) &= \beta_1 \cdot d.weight_A + \beta_2 \cdot d.weight_W \\
        &+ \beta_3 \cdot d.weight_R + \beta_4 \cdot d.weight_S \\
        &+ \beta_5 \cdot \#d.clusters + cons
    \end{split}
\end{equation*}

To test this regression, a hypotheses was defined as follows:
\begin{itemize}
    \item $H_0$: $\beta_1 = \beta_2 = \beta_3 = \beta_4 = \beta_5 = 0$; meaning that the complexity of a decomposition does not have a relation with any of the five parameters
    
    \item $H_1$: $\beta_1 \neq 0 \lor \beta_2 \neq 0 \lor \beta_3 \neq 0 \lor \beta_4 \neq 0 \lor \beta_5 \neq 0$; meaning that the complexity of a decomposition does have a relation with at least one of the five parameters 
\end{itemize}

\begin{table*}[t]
\centering
\normalsize
{\small
\caption{Comparison of the impact of similarity measures on complexity for both analysis on LdoD}
\label{tab:SimMeasuresVsComplexityOnLdoD}
    \begin{tabular}{c|c|c|c|c|c|c|}
        \cline{2-7}
        & \multicolumn{2}{c|}{Static analysis} & \multicolumn{2}{c|}{Tests}  & \multicolumn{2}{c|}{Simulation} \\ 
        \cline{2-7} 
        & Coef. & 95\% Interval & Coef. & 95\% Interval & Coef. & 95\% Interval \\ \hline
        \multicolumn{1}{|c|}{N} & 0.0230 & {[}0.021, 0.025{]} & 0.0253 & {[}0.023, 0.028{]}  & 0.0206 & {[}0.019, 0.023{]} \\ \hline
        \multicolumn{1}{|c|}{A} & 0.0035 & {[}0.003, 0.004{]} & -0.0003 & {[}-0.001, -9.14e-05{]} & 0.0017 & {[}0.002, 0.002{]} \\ \hline
        \multicolumn{1}{|c|}{W} & 0.0041 & {[}0.004, 0.004{]} & 2.781e-05 & {[}-0.000, 0.000{]} & 0.0079 & {[}0.008, 0.008{]} \\ \hline
        \multicolumn{1}{|c|}{R} & 0.0039 & {[}0.004, 0.004{]} & -0.0002 & {[}-0.000, 7.17e-05{]} & 0.0007 & {[}0.001, 0.001{]} \\ \hline
        \multicolumn{1}{|c|}{S} & -0.0002 & {[}-0.000,4.66e-05{]} & 0.0002 & {[}-2.19e-05, 0.000{]} & 0.0018 & {[}0.002, 0.002{]} \\ \hline
        \multicolumn{1}{|c|}{$R^2$} & \multicolumn{2}{c|}{0.434} & \multicolumn{2}{c|}{0.176} & \multicolumn{2}{c|}{0.682} \\ \hline
    \end{tabular}
    \vspace{-5mm}
}
\end{table*}

\begin{table}[t]
\centering
\normalsize
{\small
\caption{Comparison of the impact of similarity measures on complexity for both analysis on BW}
\label{tab:SimMeasuresVsComplexityOnBW}
    \begin{tabular}{c|c|c|c|c|}
        \cline{2-5}
        & \multicolumn{2}{c|}{Static analysis} & \multicolumn{2}{c|}{Simulation} \\ \cline{2-5} 
        & Coef. & 95\% Interval & Coef. & 95\% Interval \\ \hline
        \multicolumn{1}{|c|}{N} & 0.0439 & {[}0.043, 0.045{]} & 0.0277 & {[}0.026, 0.029{]} \\ \hline
        \multicolumn{1}{|c|}{A} & 0.0014 & {[}0.001, 0.002{]} & -0.0011 & {[}-0.001, -0.001{]} \\ \hline
        \multicolumn{1}{|c|}{W} & 0.0019 & {[}0.002, 0.002{]} & 0.0002 & {[}-9.17e-08, 0.000{]} \\ \hline
        \multicolumn{1}{|c|}{R} & 0.0016 & {[}0.001, 0.002{]} & -0.0013 & {[}-0.001, -0.001{]} \\ \hline
        \multicolumn{1}{|c|}{S} & 0.0021 & {[}0.002, 0.002{]} & 0.0019 & {[}0.002, 0.002{]} \\ \hline
        \multicolumn{1}{|c|}{$R^2$} & \multicolumn{2}{c|}{0.632} & \multicolumn{2}{c|}{0.476} \\ \hline
    \vspace{-2mm}
    \end{tabular}
}
\end{table}

The results for systems LdoD and BW are presented in Tables~\ref{tab:SimMeasuresVsComplexityOnLdoD} and \ref{tab:SimMeasuresVsComplexityOnBW}. The regression results concerning the impact of the combination of the similarity measures and number of clusters on the complexity metric show that the dynamic and static analysis have statistically significant positive correlation with complexity for the coefficients of the number of clusters. 

Regarding the similarity measures, all the analysis show that, independently of using statically or dynamically collected data, it is not possible to infer that one similarity measure by itself is determinant to generate a decomposition with the lowest complexity, because the magnitude of the coefficients is not pronounced and some confidence intervals contain the zero.

The obtained $R^{2}$ values were considerably high with the exception of functional testing environment in system LdoD with just 0.176. This means that, apart from this specific environment, the regression model explains most of the data-set (low variability).

\subsection{Best Complexity Decomposition}

Although, it seems that both collection techniques provide similar insight in terms of the correlation between the similarity measures and the complexity metric, we want to know whether they produce significantly different decompositions. 

To assess the results of the two techniques, we compare the highest quality decompositions, in terms of complexity, from each approach with a decomposition proposed by a domain expert, for both systems. In this analysis we consider the expert decompositions as reference and evaluate, using the MoJoFM metric, which approach provides closer results to it.
Since the two techniques may miss some domain entities during the collection phase, we decided that all the unassigned entities would be put in the biggest cluster, as this strategy conforms with the incremental decomposition strategy rationale~\cite[Chapter~13]{richardsonmicroservices}.

\begin{table}[htp]
\centering
\caption{Comparing generated with expert decompositions}
\begin{tabular}{cc|c|c|c?l|c|}
\cline{3-7}
\multicolumn{1}{l}{}                     & \multicolumn{1}{l|}{} & \multicolumn{3}{c?}{LdoD}            & \multicolumn{2}{c|}{BW}                  \\ \cline{3-7} 
                                         &                       & Static   & Tests    & Sim   & Static   & \multicolumn{1}{l|}{Sim} \\ \hline
\multicolumn{1}{|c|}{\multirow{8}{*}{N}} & 3                     & 62.12  & 65.15  & 68.18      & 46.67  & 44.44                          \\ \cline{2-7} 
\multicolumn{1}{|c|}{}                   & 4                     & 60.61  & 69.7   & 66.67      & 44.44  & 46.67                              \\ \cline{2-7} 
\multicolumn{1}{|c|}{}                   & 5                     & 56.06  & 68.18  & 66.67      & 44.44  & 60.00                           \\ \cline{2-7} 
\multicolumn{1}{|c|}{}                   & 6                     & 78.79  & 66.67  & 66.67      & 62.22  & 57.78                           \\ \cline{2-7} 
\multicolumn{1}{|c|}{}                   & 7                     & 77.27  & 74.24  & 68.18      & 66.67  & 64.44                          \\ \cline{2-7} 
\multicolumn{1}{|c|}{}                   & 8                     & 83.33  & 72.73  & 59.09      & 66.67  & 62.22                           \\ \cline{2-7} 
\multicolumn{1}{|c|}{}                   & 9                     & 81.82  & 74.24  & 57.58      & 71.11  & 62.22                           \\ \cline{2-7} 
\multicolumn{1}{|c|}{}                   & 10                    & 45.45  & 74.24  & 56.06      & 71.11  & 62.22                           \\ \hline
\multicolumn{2}{|c|}{avg}                                        & 68.18  & 70.64  & 63.64      & 59.17  & 57.5                           \\ \hline
\end{tabular}
\label{tab:mojoStaticAllVsDynamicAll}
\end{table}

The results from the comparisons are represented in Table~\ref{tab:mojoStaticAllVsDynamicAll}, where each cell indicates the MoJoFM percentage value (0 - 100\%) between the lowest complexity decomposition with N clusters, using a particular collection technique, and the system's expert decomposition. Overall, the MoJoFM values obtained for the different collection approaches were very similar, for both systems, which leads us to conclude that there isn't a collection technique that provides better results. However, note that, especially on the simulation technique, the dynamic analysis didn't cover all controllers during the collection phase and also missed more entities than the static approach, see Tables~\ref{tab:coveredControllersAndCoveredEntities} and ~\ref{tab:coverageControllersAndCoveredEntitiesPerController}. Therefore, we decided to assess if the dynamic analysis approach could surpass the static analysis if only the common controllers and entities were considered.

To evaluate this scenario, we re-ran the static analysis on the two monoliths considering only the common controllers and domain entities, for each dynamic technique. The results are represented in Table~\ref{tab:mojoStaticLoweredAndDynamicLoweredVSExpert}, where we can observe that, on average, both approaches continue to generate decompositions almost equally distant to the expert's, for both systems.
The major noticed difference (7-9\%), for system LdoD, is the average MoJoFM values obtained for the static approach when \textit{evened} with the dynamic analysis using the expert simulation approach. However, a similar impact is not seen for system BW.

\begin{table}[htp]
\centering
\caption{Comparing generated with expert decompositions, considering only the common controllers and entities}
\begin{tabular}{cc|c|c?c|c?c|c|}
\cline{3-8}
\multicolumn{1}{l}{}                     & \multicolumn{1}{l|}{} & \multicolumn{4}{c?}{LdoD}                       & \multicolumn{2}{c|}{BW} \\ \cline{3-8} 
\multicolumn{1}{l}{}                     & \multicolumn{1}{l|}{} & Static   & Tests    & Static   & Sim          & Static     & Sim   \\ \hline
\multicolumn{1}{|c|}{\multirow{8}{*}{N}} & 3                     & 65.15  & 59.09  & 63.64  & 71.21      & 46.67    & 44.44        \\ \cline{2-8} 
\multicolumn{1}{|c|}{}                   & 4                     & 51.52  & 69.7   & 62.12  & 71.21      & 51.11    & 46.67           \\ \cline{2-8} 
\multicolumn{1}{|c|}{}                   & 5                     & 72.73  & 68.18  & 63.64  & 66.67      & 53.33    & 60.00        \\ \cline{2-8} 
\multicolumn{1}{|c|}{}                   & 6                     & 72.73  & 54.55  & 68.18  & 66.67      & 51.11    & 57.78       \\ \cline{2-8} 
\multicolumn{1}{|c|}{}                   & 7                     & 75.76  & 74.24  & 63.64  & 69.7       & 68.89    & 64.44        \\ \cline{2-8} 
\multicolumn{1}{|c|}{}                   & 8                     & 74.24  & 72.73  & 68.18  & 59.09      & 66.67    & 62.22        \\ \cline{2-8} 
\multicolumn{1}{|c|}{}                   & 9                     & 72.73  & 74.24  & 57.58  & 57.58      & 68.89    & 62.22        \\ \cline{2-8} 
\multicolumn{1}{|c|}{}                   & 10                    & 68.18  & 72.73  & 56.06  & 56.06      & 77.78    & 62.22        \\ \hline
\multicolumn{2}{|c|}{avg}                                        & 69.13  & 68.18  & 62.88  & 64.77      & 60.56    & 57.5        \\ \hline
\end{tabular}
\label{tab:mojoStaticLoweredAndDynamicLoweredVSExpert}
\end{table}

Based on these results, we conclude that, for both systems, we don't see significant differences between the lowest complexity decompositions obtained using statically and dynamically collected data, and that none of the approaches achieve identical decompositions to the expert's, since the average MoJoFM values obtained vary around 60-70\%.

Given the similarities when compared to the expert, we assessed how far apart the static and dynamic decompositions were from each other, considering the common controllers and entities.

\begin{table}[htp]
\centering
\caption{Comparing static with dynamic decompositions, considering only the common controllers and entities}
\begin{tabular}{cc|c|l|c|l|c|l|}
\cline{3-8}
\multicolumn{1}{l}{} & \multicolumn{1}{l|}{} & \multicolumn{4}{c?}{LdoD} & \multicolumn{2}{c|}{BW} \\ \cline{3-8} 
\multicolumn{1}{l}{} & \multicolumn{1}{l|}{} & \multicolumn{2}{c?}{Static vs Tests} & \multicolumn{2}{c?}{Static vs Sim} & \multicolumn{2}{c|}{Static vs Sim} \\ \hline
\multicolumn{1}{|c|}{\multirow{8}{*}{N}} & 3 & \multicolumn{2}{c?}{57.41} & \multicolumn{2}{c?}{80.77} & \multicolumn{2}{c|}{83.33} \\ \cline{2-8} 
\multicolumn{1}{|c|}{}                   & 4 & \multicolumn{2}{c?}{83.02} & \multicolumn{2}{c?}{82.35} & \multicolumn{2}{c|}{63.41} \\ \cline{2-8} 
\multicolumn{1}{|c|}{}                   & 5 & \multicolumn{2}{c?}{78.85} & \multicolumn{2}{c?}{80.39} & \multicolumn{2}{c|}{50.00} \\ \cline{2-8} 
\multicolumn{1}{|c|}{}                   & 6 & \multicolumn{2}{c?}{78.85} & \multicolumn{2}{c?}{78.00} & \multicolumn{2}{c|}{58.97} \\ \cline{2-8} 
\multicolumn{1}{|c|}{}                   & 7 & \multicolumn{2}{c?}{78.85} & \multicolumn{2}{c?}{74.00} & \multicolumn{2}{c|}{57.89} \\ \cline{2-8} 
\multicolumn{1}{|c|}{}                   & 8 & \multicolumn{2}{c?}{80.77} & \multicolumn{2}{c?}{61.22} & \multicolumn{2}{c|}{50.00} \\ \cline{2-8} 
\multicolumn{1}{|c|}{}                   & 9 & \multicolumn{2}{c?}{78.85} & \multicolumn{2}{c?}{48.98} & \multicolumn{2}{c|}{50.00} \\ \cline{2-8} 
\multicolumn{1}{|c|}{}                  & 10 & \multicolumn{2}{c?}{60.00} & \multicolumn{2}{c?}{46.94} & \multicolumn{2}{c|}{37.84} \\ \hline
\multicolumn{2}{|c|}{avg}                    & \multicolumn{2}{c?}{74.58} & \multicolumn{2}{c?}{69.08} & \multicolumn{2}{c|}{56.43} \\ \hline
\end{tabular}
\label{tab:mojoStaticLoweredVsDynamicLowered}
\end{table}

Table~\ref{tab:mojoStaticLoweredVsDynamicLowered} presents the results of applying the MoJo metric to the best decompositions of LdoD and BW. For LdoD, the average MoJoFM between the evened static and tests approaches is 75\%, while between the evened static and simulation approaches is 69\%. For BW, the average MoJoFM between the evened static and simulation approaches was 56\%. Therefore, we can observe that the best decompositions generated by the collection techniques tend to be closer to each other than to the expert decomposition for monolith LdoD. However, the same conclusion cannot be drawn for monolith BW.

We have done an additional analysis, by inspecting the best decomposition for each one of the evened techniques, and we could observe that the clusters in the experts decomposition were more balanced in terms of the number of domain entities per cluster. This may be an indication that the expert cut was driven by the structural qualities of the monolith, which drive the domain model design. Anyway, when comparing the generated decompositions we found similarities between the semantics of the clusters.

Overall, this suggests that neither of the analysis techniques outperforms the other, even though there is space for future research.

\section{Related Work}
\label{section:related}

In recent years, a myriad of approaches to support the migration of monolith systems to microservices architectures have been proposed~\cite{Ahmadvand16,Gysel16,Hassan16,Baresi17,Klock17,Mazlami17,Fritzsch18,Nakazawa18,Lauretis19,Barbosa20,Selmadji20,Zhang20}, which use the monolith specification, codebase, services interfaces, runtime behavior, and project development data to recommend the best decompositions~\cite{Ponce19}.

In this paper we address the approaches that use the monolith codebase or runtime behavior. Although they follow the same steps, they diverge on what is their main concern and, consequently, on the similarity measures that they use, such as accesses~\cite{Jin19}, reads~\cite{Amiri18, Tyszberowicz18}, writes~\cite{Amiri18, Tyszberowicz18}, and sequences~\cite{Amiri18}. On the other hand, some authors use execution traces to collect the behavior of the monolith, e.g.~\cite{Jin19,Zhang20}, but there is no empirical evidence on whether it provides better data than the static mechanisms, and what is the required effort to collect the data, although the problem of analysing a large amount of data was already reported in a another context~\cite{Cornelissen07}. Runtime traces are used in~\cite{Eyitemi20} to calculate the percentage of calls between packages to identify a microservices decomposition, but they do not discuss the completeness of the data collection. As far as our knowledge goes, there is no work on the comparison between the use of static and dynamic analysis in the migration of monolith systems to a microservices architectures. 

Some of approaches also use different metrics to assess the result of their decompositions. Therefore, we studied the literature on microservices quality to identify which metrics to consider. The metric we used for evaluating the complexity of the decompositions are based on current state of the art metrics for service-oriented systems~\cite{Bogner17}. We applied the complexity metric for the migration of monolith systems to microservices architecture~\cite{Santos20}, which was extended to also consider several traces for a functionality, due to the result of the dynamic collection the data. Other complexity metrics use the percentage of services with support for transactions~\cite{Hirzalla09}, but they lack an integrated perspective that we provide by defining the transactional complexity of a functionality. Another complexity metric considers the number of operations and services that can be executed in response to an incoming request~\cite{Perepletchikov07}, while we consider the complexity of implementing a local transaction in the terms of inter-functionalities interactions, which emphasizes the complexity of cognitive load, i.e., the total number of other functionalities to consider when redesigning a functionality.

There is work that integrates static and dynamic analysis. For instance, in~\cite{Patel09}, static analysis is used to complement the incompleteness of dynamic analysis, in order to increase programming comprehensibility. Recent work on the migration of microservices also integrates static and dynamic analysis techniques~\cite{Krause20, Matias20}, by complementing the data collected through static analysis with dynamic analysis collected data. None of these approaches evaluates or discusses the quality of data obtained with each one of the techniques.

\section{Discussion}
\label{section:discussion}

\subsection{Lessons Learned}

From this research we learned the following lessons:

\begin{itemize}
    \item It is not possible to conclude that the decompositions generated using one of the analysis techniques, static or dynamic, outperforms the other.
    \item The effort to collect data dynamically is significantly superior than the static collection, specially when collecting and evaluating data from production, which resulted in a large amount of collected data and a very low coverage. On the other hand, the use of integration tests, that achieved better coverage, has a high development cost, because, contrary to unit tests, which aim to have 100\% coverage, integration tests, which are harder to develop and maintain, are usually designed to verify the modules integration, not the execution of all paths.
\end{itemize}

\subsection{Threats to Validity}

\subsubsection{Internal Validity}


Since dynamic analysis adds an extra layer of computation on top of the monitored systems runtime behaviour, the assumptions made on the instrumentation, to minimize the performance degradation perceived by end-users, can biase the obtained results given that: (i) an iterable object type is considered to be the type of the first element and (ii) new records are discarded when Kieker's queue is full. Concerning (i) it is somehow balanced by the fact that the static analysis may also not identify the types of objects due to dynamic binding. In what regards (ii), in the collection done through tests and simulation the probability of this situation to occur is low, because it is a single user and the amount of data in the database is small.

The approach of placing the entities not found during the collection process into the biggest cluster, when comparing the static and dynamic decompositions with the expert's, may have biased our results, as there is a probability associated with the expert decomposition that may or may not contain those entities in the same cluster. However, we also made the comparisons using other approaches and achieved similar results, thus, we are confident in discarding this as a threat.

\subsubsection{External Validity}


Due to the effort associated with the dynamic collection of data, we only analyzed two systems, but from the comparison with the decompositions generated from statically collected data, we may extrapolate that the quality of one decomposition does not outperforms the other, though the dynamic analysis of more monoliths is necessary. Nevertheless, the conclusions about the incompleteness of data and required effort associated with the dynamic collection of data are evident and shows that a cost/benefit relation may tend for the static analysis approaches.

Due to the diversity of metrics that exist for complexity can our results be generalized? We have done an analysis of the state of the art on metrics for microservices. Despite this diversity, we are confident that the results are relevant because the several metrics analyse the same elements. Our complexity metric focus on the complexity introduced by transactions and the complexity of the interactions, like other metrics do.

As described in the related work, several similarity measures have been defined to feed the automatic decomposition algorithms. In this works we have focused on the measures that correlated domain entities access, which cover a significant number of the existing approaches.

\subsection{Future Work}

As a consequence of the results of this research and the learned lessons we identify the following topics for future work:

\begin{itemize}
    \item Further explore the results of the dynamic collection of data, in terms of the frequency of each of the functionalities, and define new similarity measures to verify if it can generate better decompositions;
    \item Investigate other sequence compression algorithms with the purpose of decreasing the JSON file size and also the time taken to process it.
\end{itemize}

\section{Conclusions}
\label{section:conclusions}
The migration of monolith systems to the microservices architecture is a complex problem that software development teams have to address when systems become more complex and larger in scale. Therefore, it is necessary to develop the methods and tools that help and guide them on the migration process. One of the most challenging problems is the identification of microservices. Several approaches have been proposed to automate such identification, which, although following the same steps, use different monolith analysis techniques, similarity measures, and metrics to evaluate the quality of the system.

In this paper, two monolith systems were analysed to study the impact of applying static and dynamic analysis on the quality of the automatically generated decompositions as well as whether a particular combination of similarity measures provides better decompositions.

As result of the experiments and analysis, we conclude that different monolith analysis techniques generate decompositions that do not outperform each other, but, it was clear that the effort required by the dynamic analysis is much superior and resulted in less coverage. Although the cost is much higher, both systems were extensively dynamically analyzed which, and compared with the static analysis, is a significant effort. 

As additional contributions, (i) the gathered data from the evaluated monolith systems, using dynamic analysis, is publicly available and can be used by third parties to do further research,
(ii) the data collectors were implemented to be as configurable and extensible as possible such that they can handle a wider variety of code bases with different JAVA technology stacks.

In terms of future work, due to the different approaches proposed to the migration of monolith systems into the microservices architecture, it is necessary to do more studies that compare static and dynamic collections of data, in the context of more systems. Additionally, this type of study needs to be extended to other variations of the approaches, besides the data collection techniques, like other similarity measures and quality metrics.

\section{Data Availability}

The data used and produced in this research is available at \url{http://doi.org/10.5281/zenodo.5675593}.

\section{Acknowledgement}
This work was supported by national funds through FCT, Fundação para a Ciência e a Tecnologia, under project
UIDB/50021/2020.

\bibliographystyle{./IEEEtran}
\bibliography{./bibliography}

\end{document}